\documentclass[10pt,conference]{IEEEtran}

\IEEEoverridecommandlockouts
\usepackage{amsmath,amssymb,amsfonts}
\usepackage[backend=biber,style=ieee,natbib=true]{biblatex} 
\usepackage{algorithmic}
\usepackage{graphicx}
\usepackage{textcomp}
\usepackage{paralist}
\usepackage{booktabs}
\usepackage{mathtools}
\usepackage{multirow}
\usepackage{xcolor}
\usepackage{makecell}
\usepackage{csquotes}
\usepackage{paralist}
\usepackage{siunitx}
\usepackage{graphicx}
\usepackage{amsmath}
\usepackage{listings}
\usepackage{mathtools}
\usepackage{tikz}

\def\BibTeX{{\rm B\kern-.05em{\sc i\kern-.025em b}\kern-.08em
    T\kern-.1667em\lower.7ex\hbox{E}\kern-.125emX}}

\newcommand\percentage[2][round-precision = 2]{%
    \SI[round-mode = places,
        scientific-notation = fixed, fixed-exponent = 0, text-series-to-math = true, propagate-math-font = true,
        output-decimal-marker={.}, #1]{#2e2}{\percent}%
}

\makeatletter
\newenvironment{btHighlight}[1][]
{\begingroup\tikzset{bt@Highlight@par/.style={#1}}\begin{lrbox}{\@tempboxa}}
{\end{lrbox}\bt@HL@box[bt@Highlight@par]{\@tempboxa}\endgroup}

\newcommand\btHL[1][]{%
  \begin{btHighlight}[#1]\bgroup\aftergroup\bt@HL@endenv%
}
\def\bt@HL@endenv{%
  \end{btHighlight}%
  \egroup
}
\newcommand{\bt@HL@box}[2][]{%
  \tikz[#1]{%
    \pgfpathrectangle{\pgfpoint{1pt}{0pt}}{\pgfpoint{\wd #2}{\ht #2}}%
    \pgfusepath{use as bounding box}%
    \node[anchor=base west, fill=orange!30,outer sep=0pt,inner xsep=1pt, inner ysep=0pt, rounded corners=3pt, minimum height=\ht\strutbox+1pt,#1]{\raisebox{1pt}{\strut}\strut\usebox{#2}};
  }%
}
\makeatother

\renewcommand{\bibfont}{\footnotesize} %
\begin{document}
\newcommand{\myparagraph}[1]{\noindent \textit{#1.}}

\lstdefinestyle{patch}{
    basicstyle=\footnotesize\ttfamily,
    moredelim=**[is][{\btHL[fill=green!30]}]{`}{`},
    moredelim=**[is][{\btHL[fill=red!30]}]{@}{@},
    escapeinside={(*}{*)},
    numbers=left,
    keywordstyle=\ttfamily\bfseries,
    stringstyle=\ttfamily\itshape
}

\newcommand*\circled[2]{\tikz[baseline=(char.base)]{
            \node[shape=circle,draw=#2,inner sep=1pt] (char) {#1};}}

\title{Bogus Bugs, Duplicates, and Revealing Comments: Data Quality Issues in NPR}

\author{\IEEEauthorblockN{Julian Aron Prenner}
\IEEEauthorblockA{\textit{Free University of Bozen-Bolzano}\\
Bozen-Bolzano, Italy \\
prenner@inf.unibz.it}
\and
\IEEEauthorblockN{Romain Robbes}
\IEEEauthorblockA{\textit{Univ. Bordeaux, CNRS, Bordeaux INP},\\ \textit{LaBRI, UMR 5800} \\
Bordeaux, France \\
romain.robbes@u-bordeaux.fr}
}

\maketitle

\begin{abstract}
The performance of a machine learning system is not only determined by the model but also, to a substantial degree, by the data it is trained on.
With the increasing use of machine learning, issues related to data quality have become a concern also in automated program repair research.
In this position paper, we report some of the data-related issues we have come across when working with several large APR datasets and benchmarks, including, for instance, duplicates or \enquote{bogus bugs}. We briefly discuss the potential impact of these problems on repair performance and propose possible remedies. We believe that more data-focused approaches could improve the performance and robustness of current and future APR systems.

\end{abstract}


\section{Introduction}
The use of machine learning is now widespread in automated program repair (APR); in particular, the employment of deep neural networks has led to the actively research subfield of neural program repair (NPR). NPR systems are data-driven, that is, they require large datasets of bugfixes. Generally, these datasets consist of parallel pairs of buggy and fixed code snippets (or methods) $(s_{buggy}$, $s_{fixed})$. A NPR model is trained to transform buggy code ($s_{buggy}$) provided as input to fixed output code ($s_{fixed}$).

In most cases, such datasets are obtained through mining commits on code platforms such as GitHub. Of course, only commits fixing a bug are eligible for inclusion in such datasets. Matching commit messages against a pre-defined set of keywords (e.g., \enquote{bug}, \enquote{issue} or \enquote{fix}) is a common method to only select relevant commits~\citep{lutellierCoCoNuTCombiningContextaware2020, tufanoEmpiricalInvestigationLearning2018, monperrusMegadiffDataset600k2021, richterTSSB3MMiningSingle2022}.

This approach is used by most current large-scale APR datasets, and, in particular by all three datasets considered in this work:
\begin{itemize}
  \item Megadiff~\citep{monperrusMegadiffDataset600k2021}, a dataset of over 600,000 Java bugs;
  \item TSSB-3M~\citep{richterTSSB3MMiningSingle2022}, a dataset of over 3 million of simple Python bugs;
  \item CoCoNuT~\citep{lutellierCoCoNuTCombiningContextaware2020}, a dataset of bugs in Java, Python, C and JavaScript containing over 20 million changes (as discussed below, it is likely that not all of these changes are actual bugfixes).
\end{itemize}

\paragraph*{Issues}
While working with above datasets we came across various issues which we would like to further discuss in this paper. 
For instance, despite a deduplication step in the creation of TSSB-3M, we found a considerable number of exact duplicates. Moreover, in all three datasets,
we find a small portion of instances that clearly do not fix a functional bug (but, for instance, only remove debugging code). We call such cases \enquote{bogus bugs} (see Figures~\ref{fig:bogus-bug-megadiff} and \ref{fig:bogus-bug-tssb}). We also find a data-related issue in the commonly used Defect4J~\cite{justDefects4JDatabaseExisting2014} APR benchmark, which we also report in this work. All of these issues are described in more detail in Section~\ref{sec:issues}.

\paragraph*{Effects on Performance and Robustness}
While the amount of problematic dataset instances is relatively low (6-7\,\%), a simple experiment suggests that bogus bugs may mask real bugs. That is, a model trained on a dataset with bogus bugs may fail to recognize and repair a real bug if input code also contains a bogus bug (which is then repaired instead of the actual bug). However, removing bogus bugs from the training data using a simple keyword-based filter and retraining the same model largely resolves this problem. We take this as possible evidence that even small data impurities can have effect on model robustness. The same experiment also shows that the degree of data impurity is likely too low to have significant effects on repair performance. We discuss this experiment and its results in more detail in Section~\ref{sec:effects}.

\paragraph*{Implications}
In light of our findings, we recommend more thorough curation of large-scale datasets. This curation must not necessarily be tedious: in many cases simple filters based on regular expressions may suffice. We would also like to see more \emph{data-centric machine learning} techniques used in NPR research. These include, apart from better data filtering, also methods such as data augmentation. We discuss implications in more detail in Section~\ref{sec:implications}. %

\section{Data Issues}\label{sec:issues}
In this section we describe several data-related issues we encountered when working with APR/NPR datasets.

\subsection{Duplicates in TSSB-3M}
We found a substantial amount of duplicates in the TSSB-3M dataset. While the \enquote{3M} in the name of this dataset indicates a size of three million bugs (3,341,232 to be precise), when removing duplicates barely a million of unique bugs is left (964,202). While investigating this issue, we found that the used deduplication method is flawed in that the commit hash as well as the repository name are used as keys for deduplication\footnote{\url{https://github.com/cedricrupb/TSSB3M/blob/8b52494ab88ac9a37f7e45ec4637e77d084e01d8/run_deduplication.py#L36}}. As the same commit may appear in several repositories, only the commit hash should have been used as a key. Doing so, reduces the size of the dataset to roughly one third (as mentioned above).

\subsection{Bogus Bugs in Megadiff and TSSB-3M}

\begin{figure}
    \centering
  \begin{lstlisting}[language=Java,style=patch,numbers=none,xleftmargin=0pt,xrightmargin=0pt,framesep=0pt, breaklines=true,postbreak=\mbox{\textcolor{red}{$\hookrightarrow$}\space},]
public void doJob(AmazonS3 s3) {
    synchronized (lock) {
        (*{\btHL[fill=red!30]System.out.println(lock);}*)
        if(lock.incrementAndGet() < n) {
            try {
                lock.wait();
            } catch (InterruptedException e) {
                Fail.fail("Should not get interrupted");
            }
\end{lstlisting}
\caption{Bogus bug in Megadiff (debug code).}
\label{fig:bogus-bug-megadiff}
\end{figure}

\begin{figure}
  \begin{lstlisting}[language=Python,style=patch,numbers=none,xleftmargin=0pt,xrightmargin=0pt,framesep=0pt, breaklines=true,postbreak=\mbox{\textcolor{red}{$\hookrightarrow$}\space},]
def process_command_line(argv):
    """
    (*{\btHL[fill=red!30]Return a 2-tuple: (settings object, args list).}*)
    (*{\btHL[fill=green!30]Return args object.}*)
    `argv` is a list of arguments, or `None` for ``sys.argv[1:]``.
    """  
\end{lstlisting}
\caption{Bogus bug in TSSB-3M (docfix).}
\label{fig:bogus-bug-tssb}
\end{figure}

While looking at dataset instances from TSSB-3M~\citep{richterTSSB3MMiningSingle2022} and Megadiff~\citep{monperrusMegadiffDataset600k2021}, we found that a small but noticeable number of bugs are not what is generally considered to be a bug or a bugfix. In particular, we observe that many of these dataset instances are related to the removal of debugging code. These include for instance deletions of print or logging statements (e.g., \texttt{System.out.println} or \texttt{Log.i}). In other cases, changes are made to documentation comments (docfixes), error messages or labels of UI elements. 
In the majority of instances, such changes likely have no functional effect on functional program behavior. We consider such cases not to be real but \emph{bogus} bugs, that is, bugs that should not be part of training or evaluation sets. Figure~\ref{fig:bogus-bug-megadiff} shows a dataset instance from Megadiff where a print statement is deleted. Similarly, in Figure~\ref{fig:bogus-bug-tssb} we see changes made to a Python docstring.

\paragraph*{Bogus bugs with debugging code}
To estimate the number of bogus bugs involving debugging code, we use a simple, case-insensitive regular expression. Only considering single-line bugs (either single line added, deleted or replaced), we match the regular expression {\footnotesize\texttt{/(?<!dia)log(?!in|on|out)|print|debug|warn/i}} against the inserted or deleted lines. This expression matches the keywords \enquote{print}, \enquote{debug}  and \enquote{warn}. We use negative look-ahead and negative look-behind to exclude keywords such as \enquote{dialog}, \enquote{login} or \enquote{logout}. Bugs with matching lines are considered bogus.

Using this simple approach, we find that at least 7.7\,\% of single-line bugs in TSSB-3M and 6.4\,\% in Megadiff are bogus bugs affecting only changes to debugging code.
Our regular expression does, of course, not have perfect precision (i.e., there may be false positives). Despite this, however, we think these numbers can be considered rough lower bounds for bogus bugs as manual inspection showed other classes of bogus bugs. For instance, manually inspecting 400 dataset instances from TSSB-3M, we found 29 changes to message strings (e.g., error or UI messages), 10 changes to URIs (including URLs), 7 docfixes, and 5 cases of version string updates (i.e., strings specifying the version of a particular software package or dependency). However, these other bug classes are difficult to match with simple keywords, making it more difficult to assess how many such cases exist in the entire datasets.

\subsection{Bogus Bugs in CoCoNuT}

The authors of CoCoNuT~\citep{lutellierCoCoNuTCombiningContextaware2020} note that in a manual inspection of 100 commits mined for the construction of their dataset 7 commits where not bug-related.
Dataset instances of CoCoNuT's dataset are not individual commits (unlike instances in TSSB-3M~\citep{richterTSSB3MMiningSingle2022} and Megadiff~\citep{monperrusMegadiffDataset600k2021}); instead, each hunk in a commit is considered as a separate dataset instance. This and the fact that the dataset seems to lack exact repository URLs (e.g. a GitHub URL) makes analysis rather tedious (commit SHA hashes are provided, however). Thus, we limit our analyses to a small random samples of 800 Java instances. 

First, we find that for 148 sampled instances (18.5\,\%), the column values for code additions and deletions were identical (ignoring whitespace). Since no changes are made, such instances are likely bogus. In 61 of these 148 cases, both column values were empty.

While no repository URLs are provided, it is still possible to associate a dataset instance with a commit by using GitHub's commit search feature. We use the GitHub API to search for a matching commits (i.e., same SHA hash). This way, we can associate the 800 sampled instances with 599 commits (some instances cannot be found and multiple instances may belong to the same commit). We further use the GitHub API to look up some metadata relating to these commits.
We find that most commits are rather complex, on average spanning 29 files (7 median) and 3111 changed lines (366 median). Given this complexity, we conjecture that many changes in the CoCoNuT dataset may in fact \emph{not} be bug related. We also consider it possible that the neural architecture proposed for CoCoNuT~\citep{lutellierCoCoNuTCombiningContextaware2020} may be stymied by data quality issues.

\subsection{Revealing Comments in Defects4J}

\begin{figure}
    \centering
\begin{lstlisting}[language=Java,style=patch,numbers=none,xleftmargin=0pt,xrightmargin=0pt,framesep=0pt, breaklines=true,postbreak=\mbox{\textcolor{red}{$\hookrightarrow$}\space},]
public static LocalDateTime fromDateFields(Date date) {
    if (date == null) {
        throw new IllegalArgumentException("The date must not be null");
    }
        (*\underline{\textbf{// handle years in era BC}}*)
    return new LocalDateTime(
        date.getYear() + 1900,
        date.getMonth() + 1,
        date.getDate(),
        date.getHours(),
        date.getMinutes(),
        date.getSeconds(),
        (((int) (date.getTime() %
    );
}
\end{lstlisting}
    \caption{Excerpt from the \emph{buggy} version of Time\#12; the comment \texttt{// handle years in era BC} was introduced as part of the patch but appears also in the buggy code. The indentation of the comment further indicates that the fix likely requires an if-statement (which is the case).  }
    \label{fig:d4j}
\end{figure}

While the above issues were found in datasets, we also encountered an issue in the Defects4J~\citep{justDefects4JDatabaseExisting2014} benchmark.
Some of the fixing code changes in Defects4J are accompanied by comments that were introduced together with the fix. Unfortunately, in Defects4J,
these comments also appear in the \emph{buggy} version of the bug. Not only do these comments give away the bug location, but they may also give neural models clues as to how to fix the bug. Only looking at the 26 bugs in \texttt{Time} project, we found 7 cases of this problem in the current version of Defects4J.
One of them (Time\#12) is shown in Figure~\ref{fig:d4j}. Here the comment \texttt{// handle years in era BC} gives away not only the bug location but it also gives a repair hint (\enquote{years in era BC}). Moreover, the indentation of the comment shows that a patch will likely involve an if statement (or less likely a loop).
It is important to note that Defects4J pre-dates current learning-based methods and originally targeted generate-and-validate APR approaches which usually ignore comments.

\section{Effects on Performance and Robustness}\label{sec:effects}
In this section we return to the bogus bugs only changing debugging code mentioned in the previous section. As stated there, we found that roughly 6\,\% of single-line dataset instances in Megadiff are bogus bugs of this kind. We want to know whether this percentage is large enough to affect
\begin{inparaenum}[i)]
    \item model \emph{performance}, that is, the number of correctly fixed dataset instances and,
    \item model \emph{robustness}, that is, if a model shows the tendency to overlook bugs in presence of debugging code (and thus fail to repair the actual bug).
\end{inparaenum}

To this end, we do the following:
\begin{itemize}
    \item We remove 30\,\% of bogus bugs that \emph{delete} debugging code from the dataset and use them to train a \emph{perturbation} model. To train this model we switch the buggy and fixed versions, that is, we use $s_{fixed}$ as input and $s_{buggy}$ as output. This means, that this perturbation model can now be used to \emph{insert} debugging code into arbitrary code snippets.
    \item The remaining data (that is, all non-bogus bugs, as well as the remaining 70\,\% of bogus bugs) is split into a training and a test set using a 9:1 ratio.
    \item Then, for each instance in the test set, we generate up to five additional perturbed test samples using the perturbation model and add them to the test set.
    \item Next, we train two repair models, one on the entire training set and one on a filtered training set with all bogus bugs (related to debugging code) removed.
    \item Finally, we evaluate both repair models on the test set (including original and perturbed bugs) by generating five patch candidates for each bug (per model). We use a very simple exact-match measure to determine whether a generated patch is correct.
\end{itemize}
All models, including the perturbation model, are seq2seq Transformers based on CodeT5~\citep{wangCodeT5IdentifierawareUnified2021}.

\begin{table}[htbp]
\caption{Repair performance of the models trained on filtered (bogus-bugs removed) and unfiltered training sets on different slices of the test set (rows).}
\begin{center}
\begin{tabular}{ccccc}
& \thead{\textbf{Top-1}\\(filtered)} & \thead{\textbf{Top-1}\\(unfiltered)} & \thead{\textbf{Top-5}\\(filtered)} & \thead{\textbf{Top-5}\\(unfiltered)} \\
\toprule
Original & \percentage{0.06799840191769876} & \textbf{\percentage{0.07479025169796244}} & \percentage{0.16156612065521375} & \textbf{\percentage{0.16628046344386735}} \\
Non-bogus (filtered) & \percentage{0.07106257874842503} & \textbf{\percentage{0.074422511549769}} & \textbf{\percentage{0.16858462830743384}} & \percentage{0.16572868542629146} \\
Perturbed & \textbf{\percentage{0.07610931531002058}} & \percentage{0.041434028798119305} & \textbf{\percentage{0.17837202468410226}} & \percentage{0.13928886276814575} \\
Perturbed (unique) & \textbf{\percentage{0.07372832920556296}} & \percentage{0.04172223280624881} & \textbf{\percentage{0.17565250523909315}} & \percentage{0.14136025909697086} \\

\bottomrule
\end{tabular}
\label{tab:results}
\end{center}
\end{table}

\subsection{Results}

In short, our results suggest that filtering out dataset instances related to debugging code changes (a type of bogus bug) can greatly increase model robustness but has little effect on repair performance.

\begin{figure}
    \centering
\begin{lstlisting}[language=Java,style=patch,numbers=none,xleftmargin=0pt,xrightmargin=0pt,framesep=0pt, breaklines=true,postbreak=\mbox{\textcolor{red}{$\hookrightarrow$}\space},]
public ScheduledReporter build(MetricRegistry registry) {
(*\circled{U}{black}\,*)@System.out.println("build: " + registry);@
  return CsvReporter.forRegistry(registry)
           .convertDurationsTo(getDurationUnit())
         (*\circled{F}{black}\,*)@.convertDurationsTo(getRateUnit())@
         (*\circled{F}{black}\,*)`.convertRatesTo(getRateUnit())`
}
\end{lstlisting}
    \caption{The two patches generated by the unfiltered model (U) and the filtered model (F) for a perturbed bug. The latter overlooks the actual bug and removes the logging statement (\texttt{System.out.println}).}
    
    \label{fig:bogus-patch}
\end{figure}

Table~\ref{tab:results} shows top-1 and top-5 performance of the two model versions on three different slices of the test set:
\begin{inparaenum}[i)]
\item the original test set (including bogus bugs),
\item the filtered test set with bogus bugs removed (non-bogus),
\item and the perturbed samples (generated by the perturbation model), that is, samples in the test set with \enquote{artificially} inserted debugging code.
\end{inparaenum}

\paragraph*{Performance}
On the original test set, the model trained on the unfiltered data performs better; this is not surprising, as the original test set contains roughly 5\,\% of bogus bugs (this percentage is now lower as some bogus bugs have been used to train the perturbation model). On the filtered test set (i.e., the original test set with samples removed that contain only changes to debugging code) both models show very similar performance. Given this results, we conclude that the 5\,\% of bogus debugging bugs in the test set do not have a significant impact on repair performance.

\paragraph*{Robustness}
Table~\ref{tab:results} also reports results for perturbed and \enquote{unique} perturbed samples. As already mentioned, we generate multiple perturbations per bug; the latter, unique subset only contains a single perturbed bug per original bug (as having multiple versions of same bug in the test set, even with slightly different debugging code, could introduce some bias). We remind the reader that these are bugs where we added an additional debugging or logging statement (using the previously mentioned dedicated perturbation model). For top-1, the model trained on filtered data correctly fixed almost twice as many bugs as the unfiltered model. For top-5 this margins decreases, but results are still clearly in favor of the filtered model. This indicates that filtering can substantially increase model robustness.

\paragraph*{Debugging code leads the unfiltered model astray}
Looking at 35 generated patches from the \enquote{perturbed} test set slice where either the filtered or unfiltered model (but not both) generated a correct patch (in a top-1 setting) we found that in 25 of them, the unfiltered model generated a patch that removed the debugging or logging statement instead of fixing the actual bug. An example of this is shown in Figure~\ref{fig:bogus-patch}, where the call to \texttt{convertDurationsTo} ought to be replaced with a call to \texttt{convertRatesTo}. The first candidate output by the filtered model was a correct patch (marked with F); the unfiltered model's patch, on the other hand, removed the call to \texttt{System.out.println} instead (marked with U). 

\paragraph*{Candidate count}
We see that the margin between the filtered and unfiltered model closes with an increasing number of fix candidates (see top-5 columns in Table~\ref{tab:results}). This is not surprising, as with each additional attempt, the probability that one of them will target the actual bug (instead of the bogus one) increases. While this indicates that model robustness can also be increased by sampling multiple fix candidates, this not only increases computational overhead but also causes more user effort when choosing a possibly correct patch out of the generated candidates (or, alternatively, puts more pressure on an NPR tool's candidate ranking or filtering subsystem).

\section{Conclusions}\label{sec:implications}
In this work we spotlighted some data issues in TSSB-3M~\citep{richterTSSB3MMiningSingle2022}, MegaDiff~\citep{monperrusMegadiffDataset600k2021}, CoCoNuT's dataset~\citep{lutellierCoCoNuTCombiningContextaware2020} and Defects4J~\citep{justDefects4JDatabaseExisting2014}. We think that our findings suggest that the topic of data quality in NPR deserves a more systematic and rigorous analysis.

\paragraph*{Simple filters can boost robustness}
Our experiment with bogus bugs related to debugging code indicates that even a small amount of noise can have a negative impact of model robustness. 
While we found no effect on model performance, we note that our filtering procedure was very simple and thus provided a reasonably good payoff (i.e., more robustness) relative to the time or effort required to implement such a filter.
Moreover, similar filters could be written for other types of bogus bugs (e.g., docfixes or bogus string changes); taken together, it may be that positive effects become visible even for model performance (our experiments were limited to debugging code bogus bugs only).

\paragraph*{Data augmentation for NPR}
At any rate, previous work in data-centric machine learning clearly shows that increasing data
quality has a beneficial impact in a variety of different tasks and modalities~\citep{northcuttConfidentLearningEstimating2022}.
For NPR, apart from better commit filtering, in particular data augmentation~\citep{shortenTextDataAugmentation2021} could be a promising and interesting technique. For instance, the generation of perturbed bugs using our perturbation model could be considered a form of data augmentation. Adding these generated samples to the training set (instead or in addition to filtering) would likely also have contributed to a better model robustness. Augmentation could also help to improve robustness in other regards. For instance, \citet{geRobustNPREvaluatingRobustnessa} found that many current NPR systems lack robustness against semantics-preserving program mutations. 

\paragraph*{A data track for the next APR challenge?}
Finally, to foster more data-centric research in the current NPR landscape it might be worth considering introducing a \enquote{data track} to the next  edition of the APR competition~\citep{shariffdeenAPRCompetition20242024}. Similar to Andrew Ng's \enquote{Data-Centric AI Competition}, in this variant of the competition, the model is fixed and participants have to submit a refined training set (e.g., with outliers removed or additional augmented instances). The (hidden) model (or even several models) are then trained on the submitted training sets and evaluated on a hidden test set. The participant whose training set obtains the best performance wins the competition.

\paragraph*{Acknowledgments}
This study has received financial support from the French State in the framework of the Investments for the Future programme IdEx université de Bordeaux.

\printbibliography{}

\end{document}